# Atmospheric turbulence profiling using multiple laser star wavefront sensors

Angela Cortés[1], Benoit Neichel[2], Andrés Guesalaga[1*], James Osborn[1], Francois Rigaut[2], Dani Guzman[1]

[1] Pontificia Universidad Católica de Chile, 4860 Vicuña Mackenna, Casilla 7820436, Santiago, Chile.
[2] Gemini Observatory Southern Operations Center, Colina el Pino s/n, Casilla 603, La Serena, Chile.

**ABSTRACT**

This paper describes the data preprocessing and reduction methods together with SLODAR analysis and wind profiling techniques for GeMS: the Gemini MCAO System.

The wavefront gradient measurements of the five GeMS's Shack-Hartmann sensors, each one pointing to a laser guide star, are combined with the DM commands sent to three deformable mirrors optically conjugated at 0, 4.5 and 9 km in order to reconstruct pseudo-open loop slopes.

These pseudo-open loop slopes are then used to reconstruct atmospheric turbulence profiles, based on the SLODAR and wind-profiling methods. We introduce the SLODAR method, and how it has been adapted to work in a close-loop, multi Laser Guide Star system. We show that our method allows characterizing the turbulence of up to 16 layers for altitudes spanning from 0 to 19 km. The data preprocessing and reduction methods are described, and results obtained from observations made in 2011 are presented. The wind profiling analysis is shown to be a powerful technique not only for characterizing the turbulence intensity, wind direction and speed, but also as it can provide a verification tool for SLODAR results. Finally, problems such as fratricide effect in multiple laser system due to Rayleigh scattering, centroid gain variations, and limitations of the method are also addressed.

## 1. INTRODUCTION

$C_n^2$ is the refractive index structure parameter that quantifies the magnitude of the atmospheric optical turbulence. Knowledge of the vertical turbulence profile is crucial to assist the tomographic process in wide-field Adaptive Optics system (Rigaut, Ellerbroek & Flicker 2000; Fusco et al 2006), as well as to perform a variety of image post-processing tasks involving point-spread function reconstruction (Britton 2006).

Additionally, information about the wind speed and direction of the turbulent layers could be used to reduce the impact of the delays present in AO systems (Poyneer, van Dam &Véran 2009).

A large fraction of current and next generation ground-based astronomical telescopes use laser guide stars (LGS) adaptive optics (AO) systems. For most of them, multiple mesospheric sodium LGSs are used.

For a system comprising several Shack-Hartmann Wave-Front Sensors (WFSs) such as the Gemini MCAO

System (GeMS) installed at the Gemini South Observatory, SLOpe Detection And Ranging (SLODAR) (Wilson 2002; Wilson, Butterley & Sarazin 2009; Butterley, Wilson & Sarazin 2006) is a method that can be used to measure the turbulence profiles. In its basic form, SLODAR estimates the relative strengths of turbulent layers at different altitudes by cross-correlating the information from two stars measured at a single WFS. The height resolution and range depend on the number of subapertures in the WFS and the angular separation between the stars.

In a multi-guide stars system the conventional SLODAR technique is modified by using the data from multiple and independent WFSs. An example of the latter can be found in Wang, Schöck & Chanan (2008) for a system using six possible baselines from four WFSs illuminated by natural guide stars.

*Email: aguesala@ing.puc.cl

We take this approach one step further by applying the technique to a system with Laser Guide Stars (LGS) that works in closed-loop. Therefore, aspects such as the cone effect, fratricide effect and pseudo-open loop slopes must be considered.

We show how the SLODAR technique can be complemented by wind profiling, i.e. time-delayed cross-correlation between different WFSs can provide additional information on layer altitudes, strengths and wind velocities.

This paper is organized as follows: in Section 2 we introduce the main characteristics of GeMS; in Section 3 we describe the SLODAR methods and how it has been adapted for the means of GeMS; Section 4 introduces the wind-profiler method and Section 5 describes on-sky results. Finally Section 6 explains how the method has been cross-calibrated and validated on a bench experiment, Section 7 assesses the limitations of the current method and Section 8 gives the conclusions.

## 2. GeMS

The main objective of this work has been the implementation of an embedded turbulence profiler in an 8-meter class telescope, and specifically in the Gemini South Multi-Conjugate Adaptive Optics System (GeMS). Figure 1 shows the main components of GeMS.

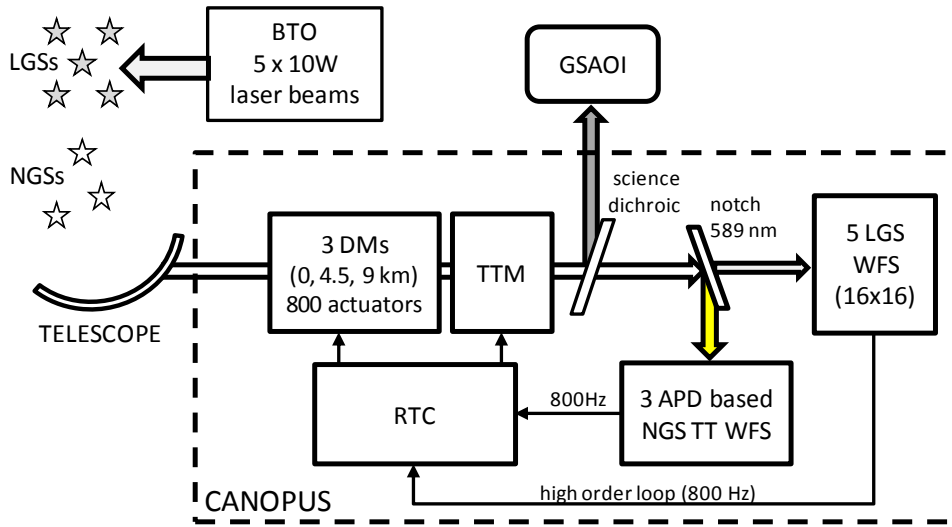

**Figure 1.** GeMS and CANOPUS

CANOPUS, the AO bench of GeMS, consists of the opto-mechanical components of the Adaptive Optics Module (AOM) and the associated sensors, mechanism, and motors. It is mounted on a side looking port of the telescope Instrument Support Structure (ISS). A flat mirror folds the beam from the telescope, which is then collimated by an off-axis parabola onto three DMs conjugated at different altitudes (0, 4.5 and 9 km respectively) and a tip-tilt mirror (TTM). The three DMs are piezo stack type, with their main parameters summarized in table 1.

**Table 1.** Main characteristics of deformable mirrors

|  | DM0 | DM4.5 | DM9 |
|---|---|---|---|
| Pitch, mm | 5 | 5 | 10 |
| Active actuators | 240 | 324 | 120 |
| Slave actuators | 53 | 92 | 88 |
| Total | 293 | 412 | 208 |

A science beam splitter transmits the infrared light to the science path, and the 589 nm wavelength from the five laser beacons are reflected by the LGS beam splitter and sent to the LGS WFS. Each WFS is a Shack-Hartmann of 16x16 with 204 valid subapertures (Fig. 2), resulting to 2040 values of slopes (axis X and Y) and working at a sampling frequency of 800 Hz (maximum). The pixel size of the WFS is about 1.38'' and the measured read out noise of 3.5 $e$. Each subaperture on the CCD uses 2x2 pixels (quadcell).

GeMS provides corrected wavefronts to the Gemini South Adaptive Optics Imager (GSAOI), an infrared camera that achieves near diffraction-limit images.

For a detailed description of this system the reader is referred to Neichel et al (2010).

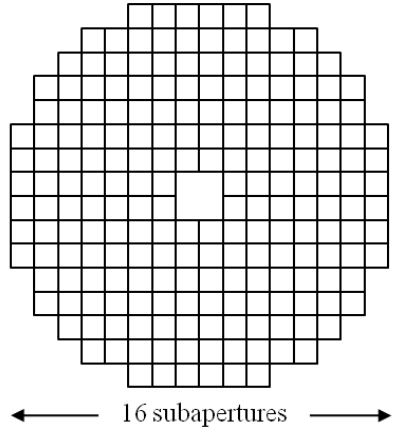

**Figure 2.** Active subapertures in GeMS's WFSs

## 3. THE SLODAR METHOD

The SLODAR method works by optical triangulation for the measurement of the atmospheric optical turbulence profile, $C_n^2(h)$, using the spatial covariance of the slopes measured by the WFSs, (phase gradient of the wavefront phase aberrations received at the ground level), each pointing at different guide stars.

The technique can estimate the turbulence strength in as many altitude bins as subapertures across the WFSs. For a WFS with $N$ subapertures across the pupil, the altitude of each layer is given by $h = md/(\theta \sec\zeta)$, where $m=\{1, .. , N-1\}$ is an integer that identifies the bin, $d$ is the size of the subaperture, $\theta$ is the relative angular separation between the NGSs and $\zeta$ is the zenith angle.

The SLODAR can be adapted to use Laser Guide Stars (LGS) as previously reported (Fusco and Costille 2010; Gilles & Ellerbroek 2010; Osborn et al 2012). In such cases, the turbulence profiling with LGS is performed to non-equally spaced bin altitudes due to the cone effect.

The altitude of these discrete bins for a pair of LGS separated by an angular distance of $\theta$ when the telescope is pointing at zenith, is given by:

$$h_m = \frac{md\,z}{z\theta + md}, \qquad (1)$$

where $z$ is the altitude of the sodium layer. Figure 3 shows the configuration for a SLODAR based on two LGS. Here, the finite distance to the stars means that the light from the guide stars form a cone. This cone effect reduces the area illuminated by the guide star at higher altitudes. According to equation (1), this effect also reduces the separation of layers ($\delta h_m$) for the higher bins. By differentiating equation (1), it can be shown that this separation or bin width can be approximated to:

$$\delta h_m = \frac{d z^2 \theta}{(z\theta + md)^2} \quad . \tag{2}$$

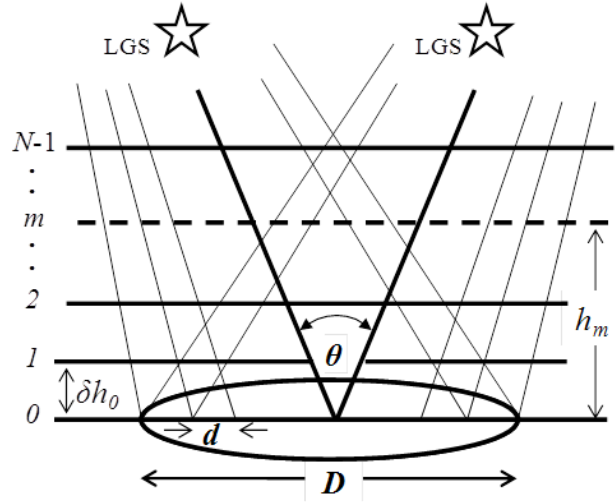

**Figure 3.** Two laser star configuration for a WFS with N x N subapertures. *D* is the telescope pupil diameter.

For a LGS asterism with a "X" shape as in GeMS (Neichel et al 2010), up to 10 pair combinations exist for the star asterism shown in Fig. 4.

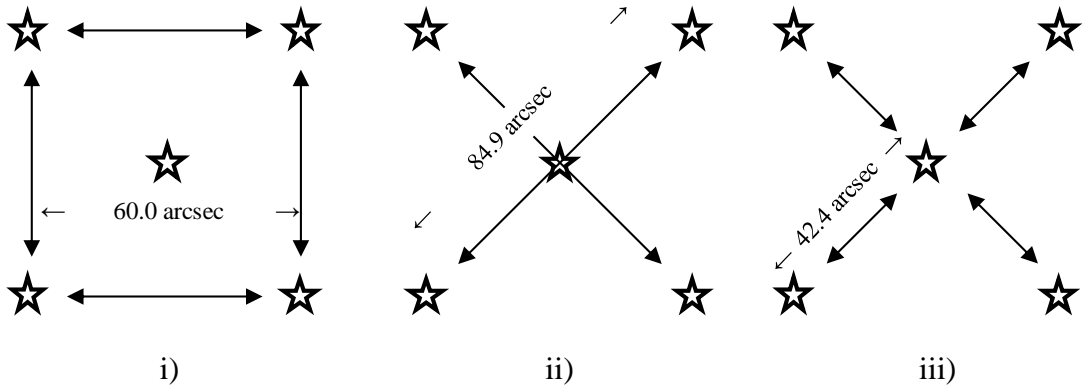

**Figure 4.** Three cases of angular separation of LGSs. Combinations i) and ii) correspond to high altitude resolution profiles whereas iii) gives low resolution and higher altitude information.

As the SLODAR altitude resolution depends on the angular separation of the guide stars, multiple altitude resolutions are possible for this configuration. For a subaperture size of $d=D/16=0.5$m and an angular separation of 60 arcseconds between the stars at the side of the asterism (the parameters used with GeMS), three cases exist:

i) Horizontal and vertical baselines between stars at the corner (4 baselines, Fig. 4 left)

$$\delta h_m = \frac{d z^2 \theta}{(z\theta + md)^2} = \frac{0.5 \cdot 60'' z^2}{(60'' z + 0.5m)^2} \quad , \tag{3}$$

ii) Diagonal baselines between stars at the corners (2 baselines, Fig. 4 center)

$$\delta h_m = \frac{\sqrt{2}dz^2\theta}{(z\theta + m\sqrt{2}d)^2} = \frac{0.5 \cdot \sqrt{2} \cdot 60" \sqrt{2}z^2}{(60"\sqrt{2} \cdot z + 0.5 \cdot \sqrt{2}m)^2} = \frac{0.5 \cdot 60" z^2}{(60" z + 0.5m)^2} \quad , \tag{4}$$

iii) Diagonal baselines between the central stars and the ones at the corners (4 baselines, Fig. 4 right)

$$\delta h_m = \frac{\sqrt{2}dz^2\theta}{(z\theta + m\sqrt{2}d)^2} = \frac{0.5 \cdot \sqrt{2} \cdot 30" \sqrt{2} \cdot z^2}{(30"\sqrt{2} \cdot z + 0.5 \cdot \sqrt{2} \cdot m)^2} = \frac{0.5 \cdot 30" z^2}{(30" z + 0.5m)^2} \quad . \tag{5}$$

Note that cases i) and ii) give the same resolution (the $\sqrt{2}$ factor appears in both the angular separation and distance between the subaperture centers). This altitude resolution is higher (smaller $\delta h_m$) than case iii).

To have an idea about the maximum altitudes that can be reached for a sodium layer at $z=90$ km, equations (1) and (2) give $h_{15}=20.04$ km with $\delta h_0=1.72$ km and $\delta h_{15}=1.04$ km for the high resolution case. In the low resolution case, the maximum altitude is $h_{15}=32.78$ km, $\delta h_0=3.44$ km and $\delta h_{15}=1.39$ km.

### 3.1. Data structure

The data collected during runs are stored in circular buffers of 24,000 frames. For the profiler we require the slopes of the AO loop residuals ($S^{res}$) and the corresponding actuator voltages ($V^{act}$).

Since GeMS normally operates in closed-loop and SLODAR requires open-loop data, it is necessary to estimate the original slopes of the incoming wavefront. This is done through the Pseudo-Open Loop (POL) reconstruction process and consists of adding the slopes of the residuals to the DMs voltages projected onto the slope domain by means of the interaction matrix (*iMat*) that corresponds to the static response of an AO system. This can be represented by the following equation:

$$S_i^{pol} = S_i^{res} + iMat * V_{i-1}^{act} \quad , \tag{6}$$

where $i$ is the discrete time. A 1-frame delay exists between the voltages and the slopes due to the exposure and readout time of the CCD.

In order to eliminate noise or low reliability subaperture slopes, the resulting slopes are subject to a masking procedure to be described later in the paper. The time-averaged centroids and piston voltages are subtracted to remove biases and bad actuators. Also, any common global motion in each WFS (Tip-Tilt) is subtracted from the corresponding slopes, so as to remove wind-shake and guiding errors. Focus removal is unnecessary, since CANOPUS compensates for sodium altitude fluctuations using a slow focus sensor (Neichel et al 2012).

The structure of vectors $S^{pol}$ and $V^{act}$ for a given frame $i$ are column vectors with $N_{ms} = 2040$ slopes and $N_{act} = 684$ voltages respectively:

$$S_i^{pol} = \begin{bmatrix} \text{slope 1, X direction, WFS1} \\ \text{slope 2, X direction, WFS1} \\ \downarrow \\ \text{slope 204, X direction, WFS1} \\ \text{slope 1, X direction, WFS2} \\ \downarrow \\ \text{slope 204, X direction, WFS5} \\ \text{slope 1, Y direction, WFS1} \\ \downarrow \\ \text{slope 204, Y direction, WFS5} \end{bmatrix} \quad ; \quad V_i^{act} = \begin{bmatrix} \text{voltage 1, DM0} \\ \downarrow \\ \text{voltage 240, DM0} \\ \text{voltage 1, DM4.5} \\ \downarrow \\ \text{voltage 324, DM4.5} \\ \text{voltage 1, DM9} \\ \downarrow \\ \text{voltage 120, DM9} \end{bmatrix} \quad . \tag{7}$$

The size of the interaction matrix is $N_{ms}$ x $N_{act}$. It reflects the effect on the measured slope when a unit control signal is applied to the corresponding actuator, i.e. it characterizes the mapping between the DMs space and the WFSs space.

## 3.2. Valid subapertures: Noise and fratricide effect

In GeMS, the lasers are launched from behind the secondary mirror and projected onto the sky. In the uplink Rayleigh scattering occurs (Fig. 5, left). This causes a contamination of the light received by the WFSs (Fig. 5, center), known as the fratricide effect (Wang, Otarola & Ellerbroek 2010; Neichel et al 2011). Another source of strong distortions is caused by vignetted subapertures along the outer ring of the WFSs.

The number of affected subapertures by these two sources of distortion can be determined by measuring the standard deviation of the slopes as clearly seen in Fig. 5 (right). It can be seen that subapertures affected by the fratricide effect present a relatively low standard deviation due to the effect of the high background on the centroid calculation (Neichel et al 2011). On the other hand, the subapertures with partial illumination present high RMS values due to the low signal-to-noise ratio in the quad-cells.

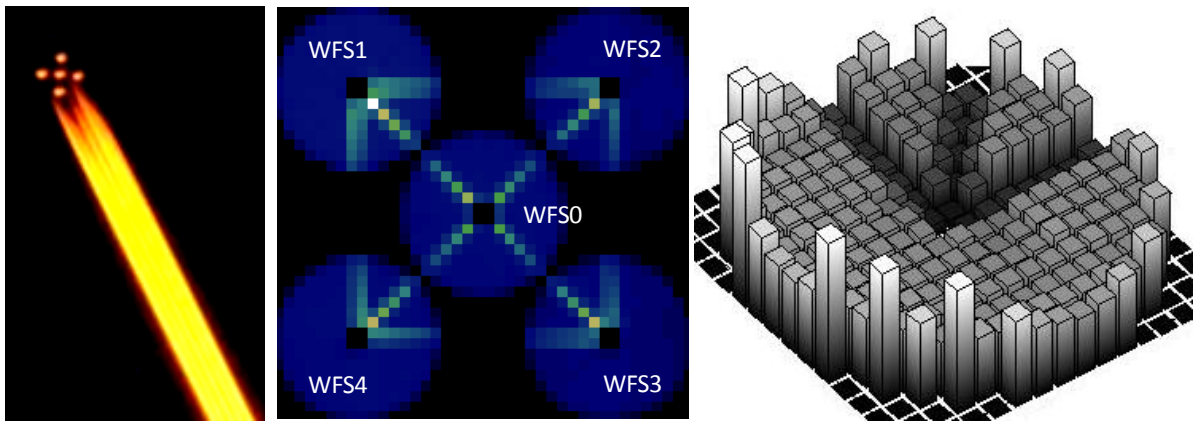

**Figure 5.** The fratricide effect. Laser projected on the sky (left) ; Photon intensity at the WFSs (center) ; RMS of slopes for WFS4 (right)

Hence, a selection process is required to eliminate the distorted subapertures. Based on the RMS values of WFS's slopes (e.g. Fig. 5, right), a mask is defined for each WFS. Examples for WFS0 and WFS1 are shown in Fig. 6.

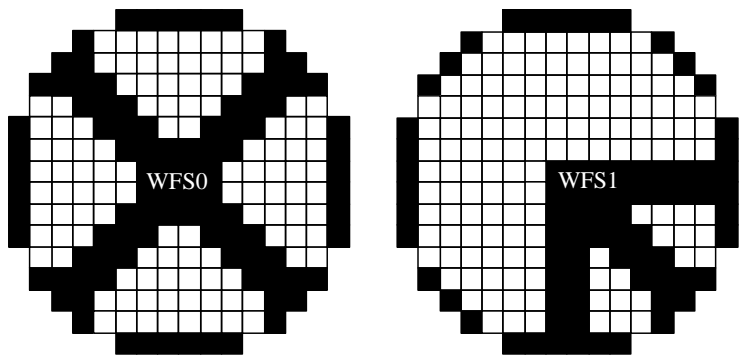

**Figure 6.** Mask used to eliminate noisy subapertures in WFS0 (left) and WFS1 (right). Unused subapertures are shown in black.

By masking some subapertures, the number of available slopes to correlate in the SLODAR method reduces substantially. Due to the elimination of the outer ring subapertures, the maximum altitude that can be measured also decreases. As table 2 shows, only 14 bins are left for the high resolution profile and 11 for the low resolution. In practical terms, the effect is even worse as the bins corresponding to the highest layers have fewer overlapping subapertures and so the signal to noise ratio is decreased. We have found that a good compromise between maximum profile altitude and noise impact is to use only the first 10 bins (13.2 km) for the high resolution profile and only 8 bins (19.0 km) for the low resolution case.

**Table 2.** Subapertures available for correlation at different altitudes. The grey cells indicate the values actually used for the profile estimation in the high and low resolution cases

| Bin | High resolution | | Low resolution | |
|---|---|---|---|---|
| | overlapping subapertures | Altitude km | overlapping subapertures | Altitude km |
| 15 | - | - | - | - |
| 14 | - | - | - | - |
| 13 | 24 | 17.9 | - | - |
| 12 | 48 | 16.8 | - | - |
| 11 | 68 | 15.6 | - | - |
| 10 | 84 | 14.4 | 8 | 24.9 |
| 9 | **108** | **13.2** | 32 | 23.0 |
| 8 | **144** | **11.9** | 64 | 21.1 |
| 7 | **164** | **10.6** | **72** | **19.0** |
| 6 | **172** | **9.3** | **72** | **16.8** |
| 5 | **252** | **7.8** | **96** | **14.4** |
| 4 | **328** | **6.4** | **136** | **11.9** |
| 3 | **396** | **4.9** | **216** | **9.3** |
| 2 | **460** | **3.3** | **272** | **6.4** |
| 1 | **536** | **1.7** | **320** | **3.3** |
| 0 | **624** | **0.0** | **368** | **0.0** |

The masking of distorted subapertures described above reduces the total number of valid POL slopes from 2040 to 1280.

### 3.3. SLODAR profiling

The optical turbulence profile is recovered by fitting theoretical impulse response functions to the cross-covariance of the centroid slope measurements. These impulse response functions express how a turbulent layer at the central bin altitudes (table 2) corresponds to the measured slope covariance functions. The response functions can be generated theoretically (Butterley, Wilson & Sarazin 2006), or in Monte–Carlo simulation. Here we use the Monte-Carlo simulation approach. We assume that the turbulence profile can be represented as a number of thin layers. We use simulated Kolmogorov phase screens positioned at the central altitude of each bin (table 2). The resulting phase at the ground for each WFS $\varphi_j^{sim}$ is transformed into slopes by simulating the WFSs of GeMS via the $D^{WFS}$ matrix:

$$S_j^{sim} = D^{WFS} \varphi_j^{sim} \quad , \tag{8}$$

where $j$ is the index of the bin.

These simulated slopes are subject to the same masking and normalization procedure carried out for the measured slopes.

To allow proper averaging of low order optical turbulence modes in the covariance matrices a large sequence of frames are required. The simulation approach follows that in reference (McGlammery 1976) with a phase-screen size of 4000 x 4000 pixels. An adequate length for the simulated sequence was found to be 10,000 frames for every bin layer. The theoretical covariance matrix for slopes in bin layer *m* is defined as:

$$C_m^{sim} = \frac{1}{10,000} \sum_{i=1}^{10,000} S_{m,i}^{sim} \cdot (S_{m,i}^{sim})^T \quad . \tag{9}$$

This set of matrices provides information about the existing correlations between slopes, however, their representation does not provide a simple understanding of such interactions. Furthermore, the size of these matrices is 2040 x 2040 so their computation, considering all layers, is highly inefficient. To reduce these problems, we use covariance maps (Vidal et al 2009). Here, the elements in the covariance matrices are rearranged and grouped taking advantage of the spatial redundancies between subapertures. This operation generates a matrix of size 330 x 330 pixels, being extremely powerful not only to reduce the computational burden, but also to understand better the following fitting process.

Figure 7 shows the result of mapping $C_6^{sim}$ (theoretical covariance with turbulence in bin 6) onto the covariance submap for bin 6 ($M_6^{sim}$). The image is first split into 4 quadrants for the X and Y slopes. This symmetrical matrix contains 10x10 submaps, each formed by the cross-correlations between pairs of WFS slopes. The diagonal groups correspond to auto-correlations for each of the 5 WFSs and the off-diagonals contain the cross-correlations of each of the WFSs slopes with every other WFS. Two characteristics are clear in the figure: i) due to a smaller optical overlapping between WFSs, the magnitude of the correlation peaks reduces for the cross-correlations in comparison with the auto-correlations; ii) The peak in the cross-correlations are displaced from their center according to the relative positions (baseline) of the correlated WFSs i.e. for the theoretical submap corresponding to bin 0, the peaks are all centered, whereas the submaps corresponding to a layer in bin 6 (for example) will have a peak displaced from the centre by 6 pixels in the direction relative to the two WFSs.

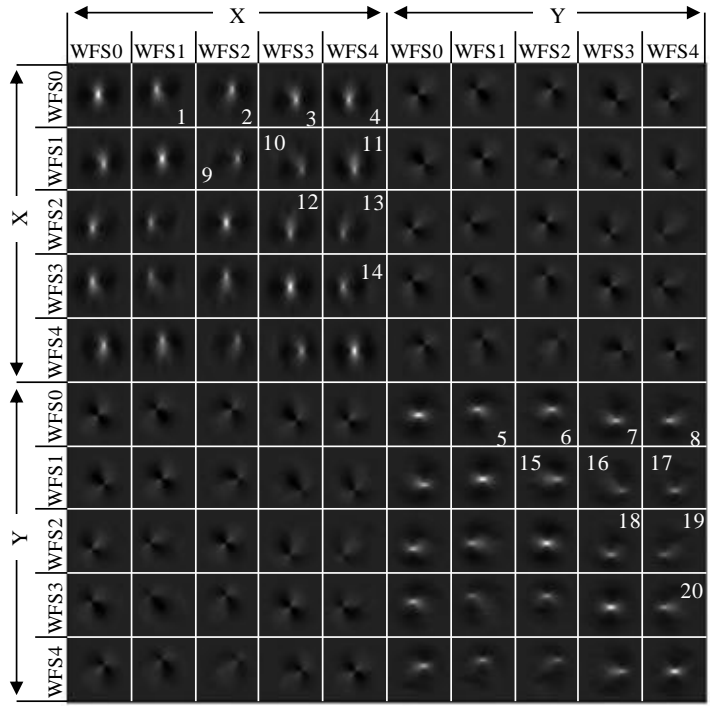

**Figure 7.** Theoretical covariance submap for $M_6^{sim}$ (bin 6). The corresponding altitudes are 9.3 km for the high resolution combinations and 16.8 km for the low resolution pairs.

The submaps of interest are those that correlate slopes in the same direction (i.e. X with X and Y with Y). Cross-correlations between X and Y slopes, are generally weaker and they are not considered in this work.

Due to symmetry there are only 20 non-redundant submaps that we use (numbered in Fig. 7). These submaps are grouped in two sets: the low altitude resolution submaps (LR, submaps 1-8) and the high altitude resolution submaps (HR, submaps 9-20). The LR submaps correspond to the shorter baselines in the asterism (case 3 in Fig. 4), i.e. the correlation between WFS0 with every other WFS (WFS$_{1-4}$) and are represented by $\mathbf{M}^{sim}_{j,LR} = \{\mathbf{M}^{sim}_{j,1},..,\mathbf{M}^{sim}_{j,8}\}$. The HR submaps correspond to the longer baselines in the asterism (case 1 and 2 in Fig. 4). They are the correlation of all possible WFS pairs, excluding WFS0 and are represented by $\mathbf{M}^{sim}_{j,HR} = \{\mathbf{M}^{sim}_{j,9},..,\mathbf{M}^{sim}_{j,20}\}$.

The diagonal submaps (auto-correlations) are used for determining the unsensed turbulence strength and noise as will be explained later in the paper.

Figure 8 shows an enlarged version of supmap 10 (covariance submap formed by the X-slopes of WFS1 and WFS3). The impulse response functions are shown for an impulse in the first bin and in bin 6. For bin 0 (ground layer), the maximum value of $\mathbf{M}^{sim}_{0,10}$ is centered. As we examine impulses in higher bins through the 3D datacube the peak of the correlation appears at increasing separations from the centre in the direction of the relative position of the WFSs.

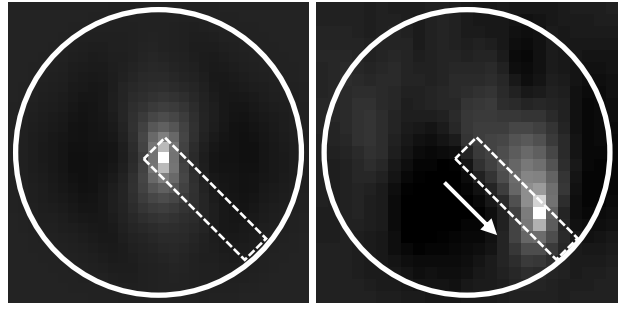

**Figure 8.** Submaps $\mathbf{M}^{sim}_{0,10}$ (left) and $\mathbf{M}^{sim}_{6,10}$ (right)

For the real data only one covariance map exists. It has the same structure of that in Fig. 7 and each submap is labelled as $\mathbf{M}^{meas}_{j}$. We then fit the impulse response functions to the real data. The resulting weights describe the contribution of each layer to the total turbulence strength measured by the WFSs, i.e. the turbulence profile,

$$\underset{\omega^{LR}}{Min} \sum_{j=1}^{8} W_j * \left[ \left( \sum_{m=0}^{L_{LR}-1} \omega^{LR}_m \mathbf{M}^{sim}_{m,j} \right) - \mathbf{M}^{meas}_{j} \right] \quad , \quad (10)$$

$$\underset{\omega^{HR}}{Min} \sum_{j=9}^{20} W_j * \left[ \left( \sum_{m=0}^{L_{HR}-1} \omega^{HR}_m \mathbf{M}^{sim}_{m,j} \right) - \mathbf{M}^{meas}_{j} \right] \quad , \quad (11)$$

where $L_{LR}$ and $L_{HR}$ are the number of bins considered for the LR and HR cases as defined in table 1, i.e. $L_{LR}$=8 and $L_{HR}$=10. In the equations above, $W_j$ corresponds to a mask for submap $j$ that selects only those values of $\mathbf{M}^{sim}_{m,j}$ and $\mathbf{M}^{meas}_{j}$ with high signal to noise ratio (the dashed rectangle in Fig. 8). Vectors $\omega^{LR}_m$ and $\omega^{HR}_m$ contain the coefficients that weigh the theoretical maps for the LR and HR cases respectively. It must be noted that the operator '*' is a matrix point-to-point multiplication.

The profile coefficients ($\omega^{LR}_m$ and $\omega^{HR}_m$) can be found using minimization methods based on gradient techniques or a matrix inversion approach based on the Truncated Least Square technique (Golub & Van Loan

1980). We found that the latter approach is faster than a gradient search, and it always converges to a global minimum. A description on how this technique is applied to the minimization problem follows.

Let us define $P_{m,j}^{sim}$ as a vector containing the result of masking theoretical covariance submap $j$ in bin $m$ as described in Fig. 8,

$$P_{m,j} = \{W_j * M_{m,j}^{sim}\} \quad . \tag{12}$$

The parenthesis {} indicates that the non-zero elements in the resulting matrix are reordered as a column vector. Vectors $P_{m,j}$ are concatenated vertically according to their submap index and the process is repeated for all simulated layers $j$ that are copied horizontally, resulting in matrices for the LR and HR cases with the form:

$$P^{LR} = \begin{bmatrix} P_{1,0} & . & . & P_{1,L_{LR}-1} \\ . & . & . & . \\ . & . & . & . \\ P_{8,0} & . & . & P_{8,L_{LR}-1} \end{bmatrix} \quad \text{and} \quad P^{HR} = \begin{bmatrix} P_{9,0} & . & . & P_{9,L_{HR}-1} \\ . & . & . & . \\ . & . & . & . \\ P_{20,0} & . & . & P_{20,L_{HR}-1} \end{bmatrix} \quad . \tag{13}$$

The next step is the construction of the measured covariance maps that follow the same path as the theoretical ones. For a given submap $j$, the masked measured covariance submap is $Q_j = \{W_j * M_j^{meas}\}$, and the full vector containing all masked submaps is

$$Q^{LR} = \begin{bmatrix} Q_1 \\ . \\ . \\ Q_8 \end{bmatrix} \quad \text{and} \quad Q^{HR} = \begin{bmatrix} Q_9 \\ . \\ . \\ Q_{20} \end{bmatrix} \quad . \tag{14}$$

Then, the contribution of each layer to the turbulence profile is

$$\omega_m^{LR} = (P^{LR})^{-1} \cdot Q^{LR} \quad \text{and} \quad \omega_m^{HR} = (P^{HR})^{-1} \cdot Q^{HR} \quad . \tag{15}$$

If no negativity constraints are imposed on the computation of the resulting profiles $\omega_m^{LR}$ and $\omega_m^{HR}$, negative values are likely to appear (Cortes et al 2011; Bendek et al 2012). These are caused by differences in the autocorrelation functions for the measured and simulated slopes as explained in detailed in Wilson et al (2009). The convenience and correctness about its use is not clear for the authors. In this work, no restrictions were imposed on the profile in which the sum of the negative values was consistently less than 3% of the total.

### 3.4. Absolute profile

The method described above will result in a relative profile. In order to obtain an absolute profile vector in terms of $C_n^2$ units, the result given by $\omega_m^{LR}$ and $\omega_m^{HR}$ requires some further processing. We know that for Kolmogorov turbulence (Hardy 1998), the Fried parameter ($r_0$) can be converted to $C_n^2$ by,

$$r_0 = \left[ 0.423 k^2 \sec \varsigma \int_0^H C_n^2(z) dz \right]^{-3/5} \quad , \tag{16}$$

The integrated turbulence over bin $m$ is $C_n^2(m) \cdot \delta h_m$. Rearranging we get,

$$C_n^2(m)\delta h_m = \frac{1}{0.423k^2 \sec\varsigma} \cdot \frac{r_0(m)^{-5/3}}{\rho_m}, \qquad (17)$$

where $\rho_m$ accounts for the stretching in $r_0$ at layer $m$ due to the cone effect. This optical spatial expansion is given by:

$$\rho_m = 1 - h_m/z \quad . \qquad (18)$$

We also know (Fried 1975) that the tilt variance integrated over a subaperture with diameter $d$ is:

$$\sigma_d^2 = 0.179\lambda^2 r_0^{-5/3} d^{-1/3} \quad . \qquad (19)$$

We now define $\sigma_0^2$ as the sum of the subaperture variances for the theoretical valid slopes, so by using the result of the minimization in equation (10) and (11), we can find a formula for the turbulence strength in bin $m$ for the LR and HR cases:

$$C_n^2(m) \cdot \delta h_m = \frac{2.37 \cdot \omega_m}{\sec\varsigma} \sigma_0^2 \quad . \qquad (20)$$

### 3.5. The unsensed turbulence

Using the SLODAR technique, it is possible to get estimates for the unsensed turbulence (turbulence above the highest bin) and also for the noise present in the measurements.

Let's define the submaps forming the diagonal of the covariance map in Fig. 7 as $\mathbf{V}^{meas} = \{V_1^{meas}, .., V_p^{meas}, ..., V_{10}^{meas}\}$, where $p$ refers to the position along the diagonal (e.g. $V_1$ and $V_6$ are the measured autocovariance submaps of WFS0 slopes in the X and Y directions respectively).

First we find the noise associated with the measurements. The central point of the auto-correlation supmaps corresponds to the centroid from each subaperture correlated with itself. As the noise is therefore correlated the central point will be equal to the slope variance plus the noise variance. As the impulse response functions are noiseless the difference between these two points (Fig. 9) can give an estimate of the noise variance. This can be estimated by,

$$\underset{\eta}{Min} \sum_{p=1}^{10} U_p * (\eta V_p^{sim} - V_p^{meas}) \quad , \qquad (21)$$

where $V_p^{sim}$ is the theoretical autocovariance submap $p$ and $U_p$ is a mask that eliminates the central point of the submaps.

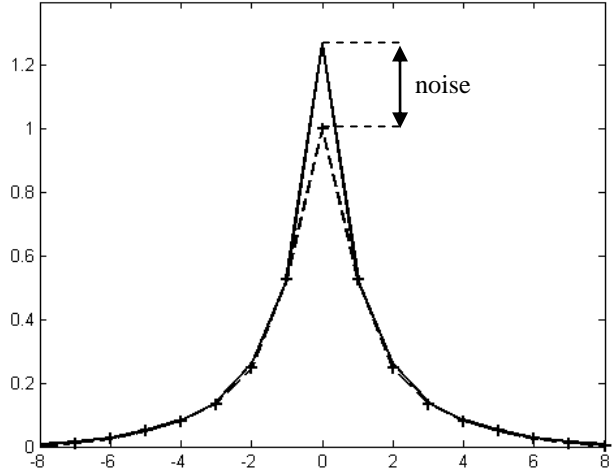

**Figure 9.** Central slice of the X-slopes submaps in the Y direction. The solid line is the measured covariance and the dotted line in the theoretical impulse function. The difference in the central pixel corresponds to noise. Data taken from on-sky observation (Tuesday, Nov 15$^{th}$, 2011, from 01:08:26 to 01:27:25).

By eliminating the noise from the measured auto covariance submaps, we can get the total turbulence above the telescope determined as the slope variance in the 5 WFS.

Finding the value of $\eta$ in equation (21) and using equation (20) we can determine the total noise-free turbulence seen by the WFSs:

$$\int_0^\infty C_n^2(h)\,dh = \frac{2.37}{\sec\zeta}\cdot\eta\cdot\sigma_0^2 \qquad . \qquad (22)$$

We now subtract the sum of turbulences from the ground up to the highest measured bin ($h_{max}$) to obtain the unsensed turbulence as:

$$C_{uns} \approx \int_{h_{max}+1}^{z} C_n^2(h)\,dh = \frac{2.37}{\sec\zeta}\cdot\eta\cdot\sigma_0^2 - \sum_m C_n^2(m)\cdot\delta h_m \quad, \qquad (23)$$

where $z$ is the sodium layer altitude. Notice that $C_{uns}$ is different for LR and HR, since $m$, $C_n^2(m)$ and $\delta h_m$ are different in each case.

## 4. WIND PROFILING

Wind profiling (Wang, Schöck & Chanan 2008) is a powerful tool that provides additional information about the turbulence structure, such as the wind velocity for each layer. It can also help to validate the results obtained via the SLODAR technique.

Using the same preprocessed data previously described for the SLODAR approach, the wind profiling consists of performing time-delayed cross-correlations between all possible combinations of wavefront measurements from the available WFSs. The method is based on the work by (Wang, Schöck & Chanan 2008) but modified to include multiple LGS-WFSs and the handling of the cone and fratricide effects.

The time-delayed cross correlation between two WFSs, WFS$_A$ and WFS$_B$, is described by the formula

$$T^{AB}(\Delta u, \Delta v, \Delta t) = \frac{\left\langle \sum_{u,v} S_{u,v}^A(t)\cdot S_{u+\Delta u, v+\Delta v}^B(t+\Delta t) \right\rangle}{O(\Delta u, \Delta v)} , \qquad (24)$$

where $S_{u,v}^{WFS}(t)$ contains the X and Y slopes of the WFS in subaperture $(u,v)$ at time $t$, $\Delta u$ and $\Delta v$ are relative subaperture displacements in the WFS grid. The time delay of the measurement, $\Delta t$, is a multiple of the acquisition time that in our case ranges from 1/800 s to 0.4 s. $\sum_{u,v}$ denotes summation over all valid overlapping illuminated subapertures, $\langle \rangle$ represents the average over the time series, and $O(\Delta u, \Delta v)$ is the number of overlapping illuminated subapertures for offset $(\Delta u, \Delta v)$.

A 2D deconvolution is applied to the time delayed cross-correlation using the simultaneous autocorrelation of each WFS using the Fast Fourier Transform (FFT), i.e.

$$FT^{-1}[FT[T^{AB}]/FT[A]] \quad , \tag{25}$$

where $A$ is the average of the autocorrelations of WFS$_A$ and WFS$_B$:

$$A(\Delta u, \Delta v) = \frac{1}{2} \frac{\left\langle \sum_{u,v} S_{u,v}^{A}(t) \cdot S_{u+\Delta u, v+\Delta v}^{A}(t) \right\rangle}{O(\Delta u, \Delta v)} + \frac{1}{2} \frac{\left\langle \sum_{u,v} S_{u,v}^{B}(t) \cdot S_{u+\Delta u, v+\Delta v}^{B}(t) \right\rangle}{O(\Delta u, \Delta v)} \quad . \tag{26}$$

When $\Delta t$ in equation (24) is zero ($T^{AB}(\Delta u, \Delta v, 0)$) the peaks along the baseline connecting the two stars represent the turbulence in the corresponding bins. This allows an alternative method for estimating the altitude of the turbulent layers. However, due to the fact that we deconvolve the covariance map with the auto-correlation (as an approximation to the altitude dependant impulse response functions) only an estimate of relative strengths can be obtained. For $\Delta t$ greater than zero, the peaks in the correlation maps move with increasing $\Delta t$ depending on the speed of wind in each turbulent layer (frozen flow hypothesis, Taylor (1938)). By tracking the movement of the peaks we can estimate the direction and speed of the wind. The peaks in the deconvolved graphs of the time cross-correlations will move according to the speed and direction of the wind in each layer, so layers too close to be resolved by the SLODAR method can be separated as long as they have different velocity vectors. It is common to find peaks at the center of the graph that do not move with increasing $\Delta t$. This is caused by the so-called dome-seeing (turbulence inside the dome of the telescope). The altitude resolution in the wind profile will depend on how well the peak is tracked, which in turn will be a function of the pixel size and the time elapsed before the frozen flow assumption breaks down.

The altitude resolution in the wind profile will depend on how well the peak is tracked, which in turn will be a function of the pixel size and the time elapsed before the frozen flow assumption breaks down. The validity of this latter assumption is essential for this technique and only a few quantitative studies have been carried out over the last two decades; mainly for predictive control (correcting for delay lags in AO loops). Gendron and Lena (1996) and Schock and Spillar (2000) carried out extensive observational campaigns and by cross-correlating WFS measurements they conclude that accurate wavefront prediction can be made under the assumption of frozen flow but only to time scales of up to 10 ms. Poyneer et al (2009) verify this hypothesis for estimation of the velocity vector variability at different layers. Their results show that for this purpose (velocity vector estimation), the atmosphere is stable for several seconds and in some cases, the wind profile had substantial stability in measurements taken 1 hour apart.

In our case the stability requirement of the atmosphere is imposed by the time evolved between two measurements where the cross-correlation peaks between two or more WFS are traceable. Determining this time is not easy, since the signal to noise ration of the peaks will depend on altitude (separation of WFS in the metapupil); wind direction (layer measured in two WFS moving to the direction of their baseline center or away from it). It will also depend on the wind strength, since for low speed layers, the correlation peaks will be hard to distinguish from the dome seeing or the dominating ground layer. On the other hand, a fast wind will mean less time available to correlate the common WFS area at higher altitudes.

In our case, data processed offline for different nights showed that correlation peaks can be tracked confidently up to 0.4 s, however, in the case of winds with speeds below 2 m/s, the peaks from different layers could not be separated. Much work remains to be done in order to determine the conditions under which the wind profiler is applicable. So far, we only limit its use to check the validity of the SLODAR estimation.

## 5. RESULTS

Using equations (20) and (23), we are able to obtain an absolute profile using the SLODAR technique.

Figure 10 presents the high and low resolution profiles obtained from on-sky data during April 15th, 2011.

The figure also shows the intensity of unsensed turbulence. As expected, the value obtained is smaller for the LR plot since the maximum height in this case is higher than in the HR case. The measured seeing is $r_0$=12.5 cm.

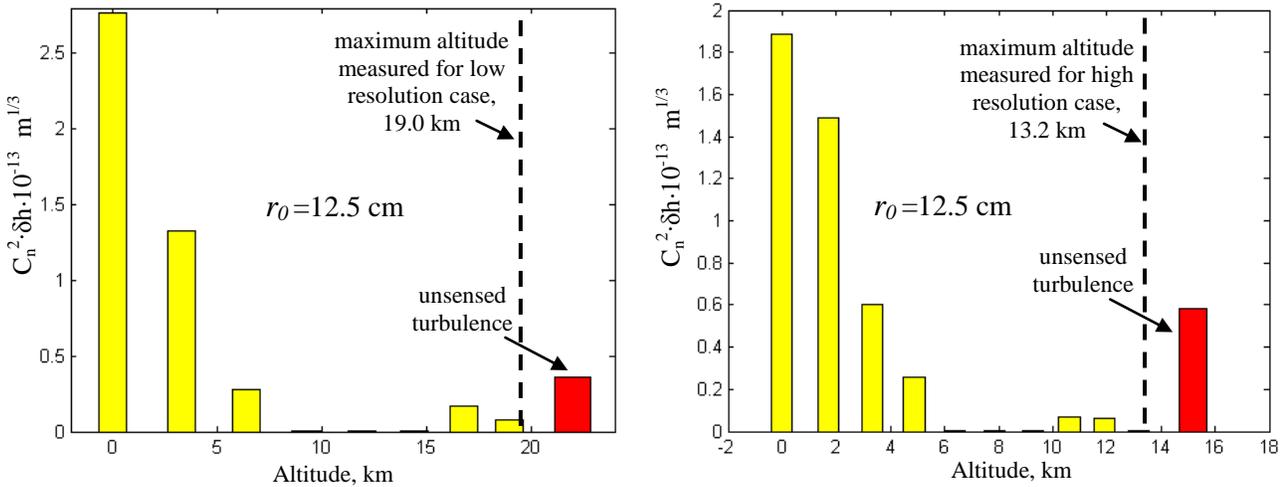

**Figure 10.** SLODAR fitting for April 15th, 2011, 23:55:15. Left: LR profile (light grey bars) and unsensed turbulence (dark grey). The broken line indicates the maximum altitude attainable. Right: HR profile.

Figure 11 shows an example of the wind profiler for the same data in the previous figure. The graph presents the average of the time-delayed correlation for the combination WFS1/WFS2 and WFS4/WFS3 (as they have the same baseline). The maximum delay is 0.5 s, but the correlation process is repeated and averaged over the entire length of the circular buffer recorded (24.000 frames at 800Hz = 5 mins). This average requires that the speed and direction of the wind remain constant for that period.

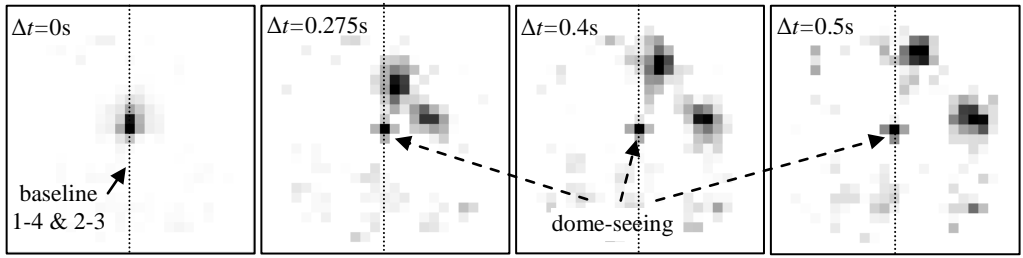

**Figure 11.** The sequence show the time correlation from $\Delta t = 0$ to $\Delta t = 0.5$s, with two layers moving in different directions and a third static peak at the center, corresponding to dome-seeing. The baseline corresponds to the vertical combinations of WFS, i.e. 1-4 and 2-3 (on-sky data, april 15th, 2011, 23:55:15).

When distinctive correlation peaks are observed in the time sequence, the speed and direction of the wind can be determined by tracking their movement. In the previous case (Fig. 11), two layers moving with vectors {8.7m/s, 86.5°} and {8.0m/s, 19.5°} are clearly distinguished, supplying extra information to the profile which is not obvious from Fig. 10 alone. A third static peak at the center of the correlation map, corresponding to dome-seeing, is also noticeable.

Figure 12 shows how this technique can also be used to determine the layer altitudes. A peak moving at high speed in the horizontal direction is observed (velocity vector: {24.1m/s, 265º}). Its origin can be traced back to the baseline where the intersection corresponds to an approximate altitude of 5.8 km. In this figure, the central pixel containing the dome-seeing and ground layer turbulence has been trimmed for a better visualization of the weaker turbulence.

In a previous article Wang, Schöck & Chanan (2008) claim that the wind profiler can provide a better resolution than SLODAR. In fact, a good tracking of the peaks can provide an accurate intersection point with the baseline where the altitude of the layer can be determined. However, there are a few issues associated with this technique. The first is that the tracking is limited by the quantization by the pixel size (equivalent to the projected subaputre size). Therefore many subapertures are required to successfully track the layers with the precision required. The second is that the frozen flow assumption holds for a limited time, so tracking the peak becomes increasingly difficult. This also means that it is difficult to get even relative layer strengths out. The third problem is that contiguous layers with different winds and strengths can coexist, making it hard to separate the individual tracks. Therefore, using both methods together further improves the validity of both results.

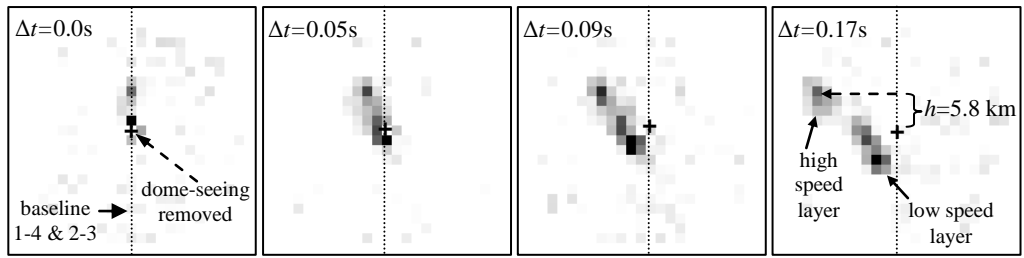

**Figure 12.** The sequence shows a fast moving layer around 5.8 km moving in the horizontal direction. The altitude is determined by intersecting the layer track and the vertical baseline. The central dome-seeing peak has been removed for better visualization (data from April 19th, 2011, 06:18:04).

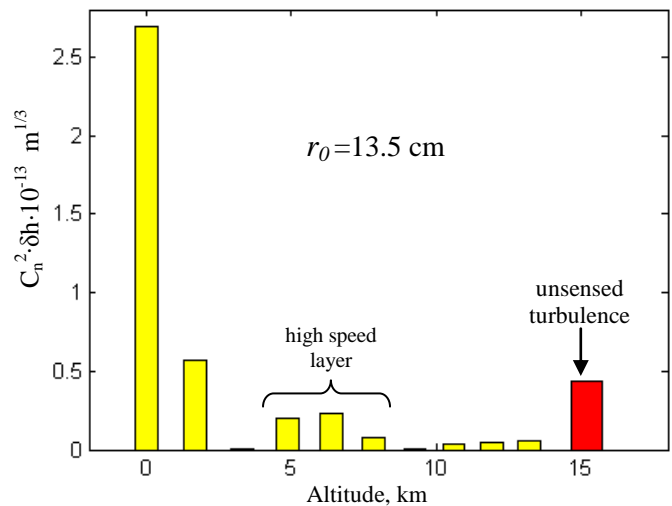

**Figure 13.** Profile observed with covariance maps. The integrated turbulence strength of each profile has been normalised to unity (on-sky data, April 19th, 2011, 06:18:04).

The high speed peak at 5.8 km confirms the existence of turbulences detected by the SLODAR as seen in Fig. 13. Here, relevant turbulence strength exists between 5 km and 8 km making it hard to track individual thin layers. The seeing measured in this case is $r_0$=13.5 cm.

Note that these two examples have been chosen randomly, and for illustration purpose only. A deeper statistical analysis of the conditions at Cerro Pachon is pending, and will be presented in a forthcoming paper. In addition, estimating the error associated to a profile and a given set of atmospheric parameters is difficult. In order to assess the main limitations of our method, we have tested it against a calibrated and controlled turbulence as presented in the next section.

## 6. CALIBRATION OF THE METHOD

Using the CANOPUS internal calibration source and the three DMs to artificially generate turbulence at 0 km, 4.5 km and 9.0 km, we tested the SLODAR and wind profiler in open and closed loop. A total of 50 runs were implemented with different turbulence settings for wind speed and direction, seeing conditions ($r_0$) and energy distribution among the three DMs. The estimated wind and turbulence parameters were compared to the ones used to generate the turbulences, getting very good agreement. As an example, Figs. 14 and 15 show the results for a turbulence generated by DM4.5 with wind velocity of 30m/s in the X direction and a $r_0$ = 42.0 cm.

The estimated values for this case were an equivalent altitude, $H_{eq}$, of 4.65 km (obtained as a weighed sum of altitudes), a wind speed of 30.8 m/s and a $r_0$ of 36.5 cm.

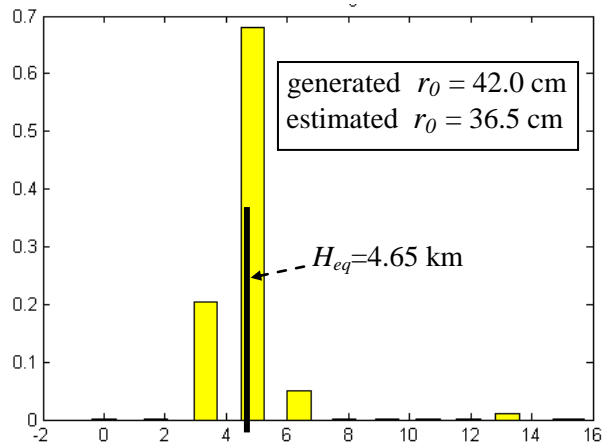

**Figure 14.** Estimated profile using the calibration source and excitation of DM at 4.5 km.

The weighed altitude estimation and wind parameters were always estimated with accuracy better than 5% for all runs. However in the case of the Fried's parameter, this was not the case and the example presented above (worst case found) errors could exceed 10%. This is thought to be caused by wrong gains in the centroid gains, that as stated in equation (6), they directly affect the errors of the POL values and hence, the measured turbulence. This implies that the centroid gain of the quad-cells become critical in the effectiveness of the method so an in-depth analysis of their error impact on the fitting accuracy is required.

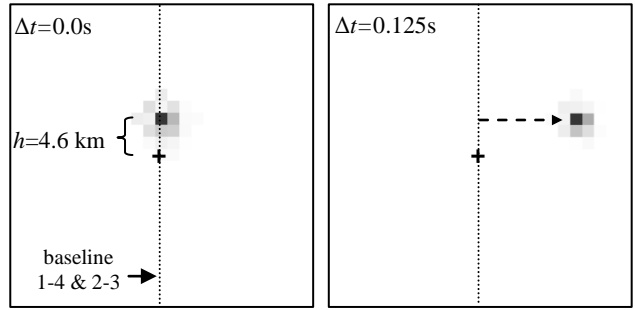

**Figure 15.** Result of wind profiler for a turbulence generated at DM4.5 with wind speed of 30 m/s in the X direction.

Data were taken from the bench in closed loop but with zero loop gain whilst applying a known turbulence on the DMs. Data in closed-loop were also taken for the same simulated turbulence and the scatter plots for two subapertures in each case are shown in Fig. 16. A noticeable difference exists in the slope gain with respect to the ideal one and also a nonlinear effect due to the quad-cell dynamic range appears at higher values of slopes amplitudes. This proved to have a low impact on the results for normalized contributions of each layer to the total turbulence strength.

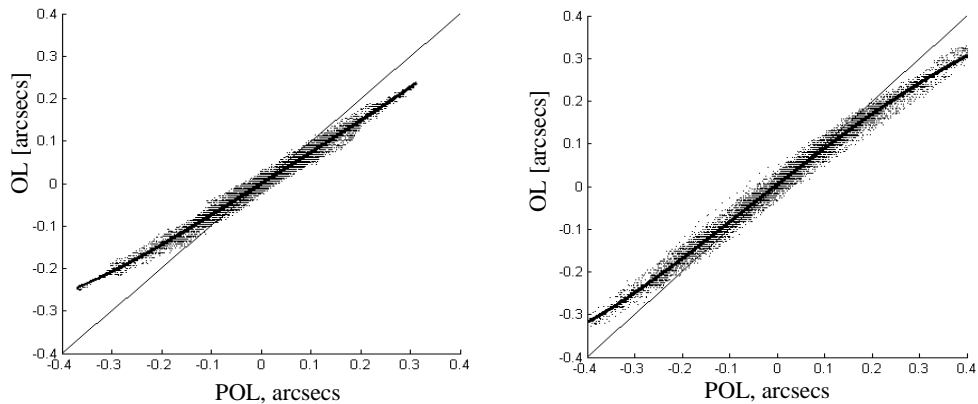

**Figure 16.** Open Loop versus Pseudo Open Loop. The thin line is the ideal relationship that should be obtained; the thick line is a 4th order polynomial fit to the data (dot cloud). Left: WFS0 subaperture 1; Right: WFS1, subaperture 57

A simulation analysis was carried out in order to quantify the effect of these mismatches on the fitting of the $C_n^2$ profile. For the high resolution case, the errors fell below 3% for deviations in the centroid gain of up to 50% respect to the correct value. For the low resolution case, however, these errors jumped to 8% for a 50% deviation.

It is interesting to note that this error had little impact for relative profiles (normalized contributions of each layer to the total turbulence strength), for both high and low resolution. Furthermore, GeMS can calibrate the centroid gains in almost real time during on-sky operation (Gratadour & Rigaut 2007), so the negative impact can be known and limited.

Simulations were also run for different values of the turbulence outer scale $L_0$, but no significant impact was found on the result. This is not surprising, since by eliminating the tip and tilt from the POL slopes, the potential effect of differences in the theoretical and measured the lower part of the spectrum are greatly attenuated.

The optimal masking to be applied to the covariance submaps shown in Fig. 8 was also a subject of further analysis via simulation and artificially generated turbulences. The best performance in both instances was found to be a masking that eliminates any submap pixel outside the line describing the baselines between WFSs.

Artificial generation of turbulence via DMs is a powerful calibration and validation tool for the method, but it must be kept in mind that the limit imposed by the actuators' pitch over the simulated turbulence spectrum, causes that the impulse response of the turbulence is different from the Kolmogorov model used for computing the theoretical submaps (Wilson 2002; Wilson, Butterley & Sarazin 2009; Butterley, Wilson & Sarazin 2006). This was clearly seen when estimating the profile using DM excitation, where negative values in some of the profile component could be equivalent to up to 2% of the total turbulence.

## 6. CONCLUSIONS

A SLODAR-based method to estimate turbulence profile embedded in a MCAO system has been described. It uses the measured slopes from five LGS WFSs in an X-asterism. Using this asterism we obtain two distinct altitude resolutions (lower resolution but higher maximum altitude using the covariance of the central guide star with each of the corners and higher altitude resolution but lower maximum altitude using the covariances between the corner guide stars along the diagonals and the edge). These altitude resolutions are altitude dependant due to the focal anisoplanatism of the laser guide stars.

As the MCAO system is closed-loop (i.e. the WFSs see the corrected wavefront) we must first estimate the pseudo open-loop slopes. This is done using an interaction matrix.

Some early results from the SLODAR profiling are shown. We find a mixture of conditions with some data showing an exponential decay of turbulence strength with altitude and others showing turbulent layers at higher altitudes. The wind profile, obtainable from the same data by calculating the cross correlation with increasing time delay of one of the guide stars, can be used to verify the SLODAR profile as the correlation peaks can be seen to move in the cross-correlations with increasing time delay. These peaks can be tracked and used to estimate altitudes and compared with the SLODAR profile. We find that the two techniques compare well. The data collected and analysed so far also shows a consistently strong dome seeing component. The concurrent wind profile can also be used to extend the turbulence profile, not only by adding verification, but also it can be used to increase the resolution as two layers very close together in altitude may fall in the same bin in the SLODAR profile but will become separated in the wind profile if they have different wind vectors. We can also use the combined information to estimate other parameters important for adaptive optics control, such as the coherence time and isoplanatic angle of the atmosphere.

We are in the process of gathering more data in order to derive statistical analysis of the atmospheric turbulence above Cerro Pachon. This work will be presented in a forthcoming paper.

## ACKNOWLEDGMENTS

This work has been supported by the Chilean Research Council (CONICYT) through grant Fondecyt 1120626, Anillo ACT-86 and scholarship for first author. We are also grateful to Andrei Tokovinin, Tim Butterley and Tim Morris for fruitful discussions.